\documentclass[prl,twocolumn,noshowpacs,preprintnumbers,amsmath,amssymb,nofootinbib,showkeys,superscriptaddress]{revtex4-2}
\usepackage{bbm}
\usepackage{lipsum}
\usepackage{mathrsfs}
\usepackage{graphicx}
\usepackage{dcolumn}
\usepackage{bm}
\usepackage{color}
\usepackage{comment}
\usepackage{times}
\usepackage{footnote}
\usepackage{txfonts}
\usepackage[hypertexnames=false]{hyperref}
\hypersetup{
    colorlinks=true,
    linkcolor=black,
    citecolor=blue,
}
\newcommand{\sectionprl}[1]{{\em #1}\/.---}

\begin{document}

\title{Anomalous Enhancement of Yield Strength due to Static Friction}

\author{Ryudo Suzuki}
\affiliation{
Department of Physics, Kyoto University, 
Kyoto 606-8502, Japan
}

\author{Takashi Matsushima}
\affiliation{
Faculty of Engineering, Information and Systems, 
University of Tsukuba, Tsukuba, Ibaraki 305-8573, Japan
}

\author{Tetsuo Yamaguchi}
\affiliation{
Department of Biomaterial Sciences, The University of 
Tokyo, 1-1-1 Yayoi, Bunkyo-ku, Tokyo 113-8657, Japan
}

\author{Marie Tani}
\affiliation{
Department of Physics, Kyoto University, 
Kyoto 606-8502, Japan
}

\author{Shin-ichi Sasa}
\affiliation{
Department of Physics, Kyoto University, 
Kyoto 606-8502, Japan
}

\date{\today}

\begin{abstract}
Friction is fundamental to mechanical stability across scales, from 
geological faults and architectural structures to granular materials 
and animal feet.
We study the mechanical stability of a minimal friction-stabilized 
structure composed of three cylindrical particles arranged in a 
triangular stack on a floor under gravity. 
We analyze the yield force, defined as the threshold compressive 
force applied quasi-statically from above at which the structure 
collapses due to sliding at the floor contact. Using singular 
perturbation analysis, we derive an expression which quantitatively 
predicts the yield force as a function of the static friction coefficient 
and a small dimensionless parameter $\epsilon$ characterizing 
elastic deformation.
\end{abstract}

\maketitle

\sectionprl{Introduction}
Friction is essential for mechanical stability across a wide 
range of systems, from geological faults \cite{Scholz2019}, 
solid blocks \cite{Perrson2013,Heslot1994}, slender or 
architected structures \cite{Duran2012}, 
granular materials \cite{Jaeger1996,Duran2012,Andreotti2013,Mitarai2006}, 
and animal feet \cite{Autumn2000,Urbakh2004}, 
down to AFM tips \cite{Socoliuc2004}.
It enables static equilibrium by resisting sliding at contacts, often in concert 
with geometric constraints. Examples include arches 
\cite{Heyman1966}, masonry domes \cite{Beatini2018}, and 
sandpiles stabilized by their angle of repose \cite{Zhou2002}, 
where friction plays a critical role in maintaining stability. 
Although the importance of friction in these systems is widely 
recognized, gaining a quantitative understanding 
of how friction contributes to mechanical stability would provide 
valuable insights into the design and analysis of friction-stabilized 
structures. 

Recent studies have highlighted that frictional systems involving 
elasticity can exhibit a rich variety of mechanical responses, 
including instabilities and transitions under applied forces 
\cite{Socoliuc2004,Heslot1994,Rubinstein2004,Yamaguchi2009,Yamaguchi2016,
Scholz1998,vanHecke2009,Liu2010,Silbert2010,Vinutha2016,Sano2017,
Tani2024,Otsuki2023SM,Guerra2021,McNamara2004}. 
A major research direction investigates how friction and elasticity 
interact to produce dynamical behaviors across scales, 
from stick-slip dynamics in microscopic contacts
\cite{Socoliuc2004} to shear-induced transitions in granular 
materials and elastic solids at laboratory scales 
\cite{Heslot1994,Rubinstein2004,Yamaguchi2009,Yamaguchi2016}, 
and further to macroscopic rupture fronts in tectonic faults and 
geophysical granular assemblies \cite{Scholz1998}. 
Another active direction focuses on the emergence of rigidity and 
stability through geometric constraints and mechanical interactions, 
particularly in the context of jamming transitions 
\cite{vanHecke2009,Liu2010}, including those involving frictional 
particles \cite{Silbert2010,Vinutha2016}.  
Remarkably, even minimal systems---consisting of a few interacting 
elements---have been shown to exhibit rich and critical 
behavior \cite{Sano2017,Tani2024,Otsuki2023SM,Guerra2021,McNamara2004}. 
These developments lead to the question of whether mechanical 
stability provided by static friction can exhibit critical behavior, 
even in the simplest system. 

\begin{figure}[t]
    \centering
    \includegraphics[width=8.8cm]{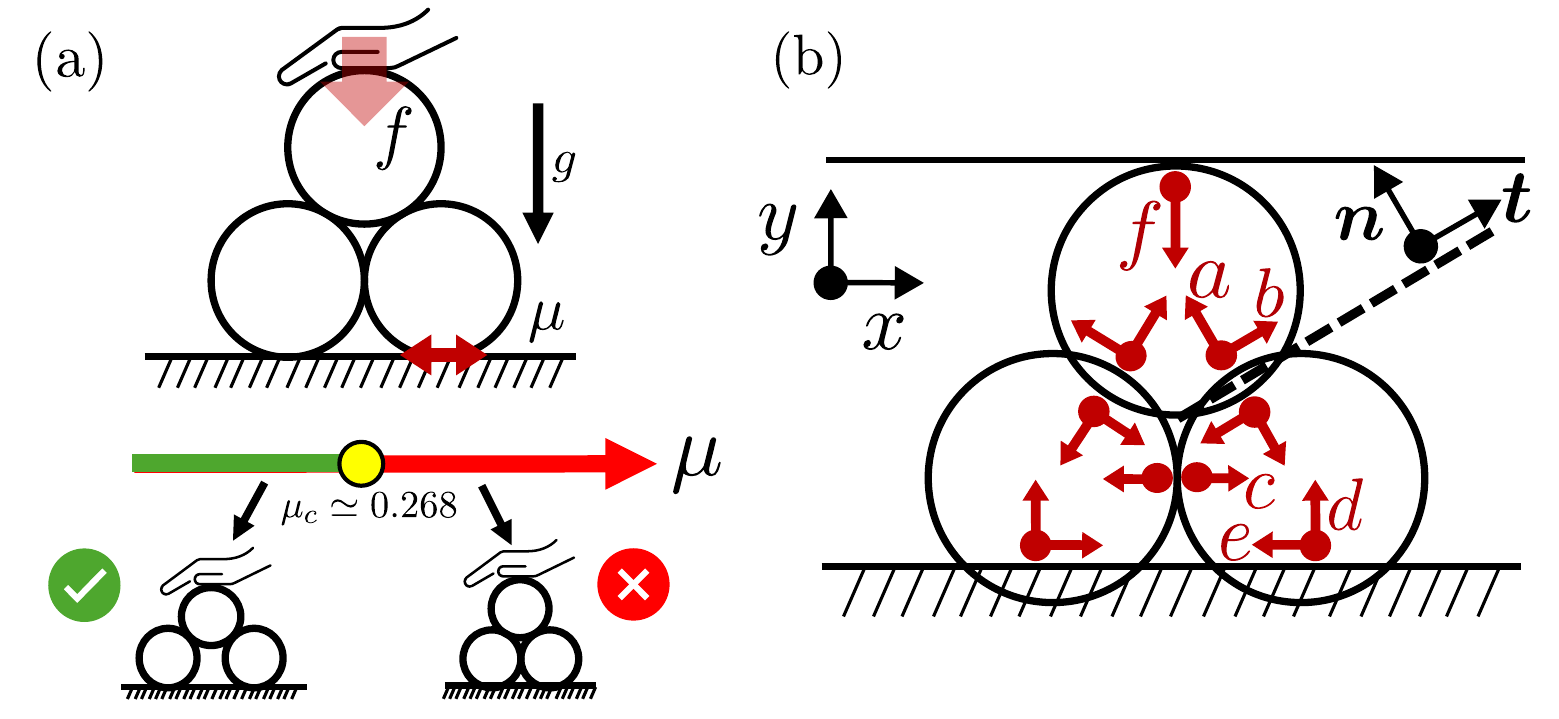}
    \caption{(a) Schematic of the system: three frictional 
    cylinders, with a cylinder-floor friction coefficient $\mu$
    are stacked under gravity and compressed from above 
    by an external force $f$ applied to the top cylinder. 
    (b) Contact forces $a,b,c,d$, and $e$ act between the 
    cylinders and the floor. 
    The $x$ and $y$ axes indicate the horizontal (along the floor) 
    and vertical (normal to the floor) directions, respectively.
    The unit normal and tangential vectors 
    $\bm{n}$ and $\bm{t}$ are defined at the contact point between cylinders.
    Force $a$ is parallel to $\bm{n}$, and $b$ is parallel to $\bm{t}$.
    Due to the law of action and reaction, forces equal in magnitude and 
    opposite in direction to those acting on the top cylinder 
    act on the bottom cylinders. 
    }
    \label{fig1}
\end{figure}

In this Letter, we study one of the simplest friction-stabilized 
systems: three identical cylindrical particles stacked under gravity 
via side-to-side contact, forming a triangular arrangement. 
A quasi-static vertical force is applied from above to the top 
cylinder. We define the yield force as the threshold above which 
the bottom cylinders slip against the floor, leading to collapse.
(See Supplemental Material for Videos S1 and S2~\cite{supplement}.)
Despite its simplicity, the system exhibits a friction-induced 
transition: in the rigid-body case, the yield force diverges 
at a critical floor friction coefficient 
$\mu_c$, which separates destructive and 
non-destructive regimes, as shown in Fig.~\ref{fig1}(a). 

We explore this transition in a realistic system where the cylinders 
can deform under load. We first perform discrete element method (DEM) 
simulations that incorporate linear elasticity and frictional contact. 
We find that the yield force exhibits singular behavior governed by 
a dimensionless stiffness parameter. We then analyze this anomalous 
enhancement of stability by performing a singular perturbation analysis. 

Our results demonstrate that mechanical stability of a piling 
structure, maintained by friction, exhibits critical scaling 
governed by elasticity and contact friction. This system 
thus provides a tractable setting that may help deepen our 
understanding of how friction contributes to stability 
in granular and other frictional assemblies.

\sectionprl{Setup}
In this Letter, we consider a system consisting of three 
frictional cylinders, each with radius $r$, axial length 
$l$, and mass density $\rho$ (mass $m = \pi \rho r^2 l$). 
They are stacked vertically on a floor with a cylinder--floor 
friction coefficient $\mu$ under gravity $g$ and compressed 
from above, as illustrated in Fig.~\ref{fig1}(a). 
The compression is applied quasi-statically to the top 
cylinder by a wall exerting a vertical force $f$.
In response, forces $a$, $b$, $c$, $d$, and $e$ arise at the 
contacts between the cylinders and between the bottom cylinders and 
the floor, as shown in Fig.~\ref{fig1}(b). Because the system is 
symmetric along the cylinder axis (i.e., along the depth of the page), 
we reduce it to a two-dimensional description by projecting 
onto the vertical plane perpendicular to the cylinder axes.

As the external force $f$ increases, the frictional force at 
the contact between the bottom cylinders and 
the floor also increases to maintain mechanical equilibrium. 
When $f$ exceeds a certain threshold, 
the frictional force reaches the maximum static friction, 
causing slip at the contact points and leading to the collapse 
of the stacked structure. This threshold value corresponds to the 
yield force introduced above.

The six variables---the contact forces $a$, $b$, $c$, $d$, and $e$, 
and the yield force $f$---are determined by the following set of 
conditions. Under quasi-static loading, the entire system remains in 
mechanical equilibrium at all times. First, from the vertical force 
balance for the entire stack, we obtain $f + 3mg = 2d$.
Next, assuming that slip occurs at the contact points between the 
bottom cylinders and the floor, the Coulomb friction law gives 
$e=\mu d$, where $\mu$ is the static friction coefficient between the 
cylinders and the floor. Additionally, we assume that the contact 
force $c$ vanishes at the onset of slip, imposing the condition 
$c=0$.

Under these assumptions, the torque and force balance equations 
for the right bottom cylinder are given by the following three equations;
those for the left one follow from symmetry.
\begin{align}
    b - e &= 0, \label{eq:elastic_balance1}\\
    a n_x + b(t_x + 1)  &= 0, \label{eq:elastic_balance2}\\
    2 a n_y + 2 b t_y - f - mg & = 0, \label{eq:elastic_balance3}
\end{align}
where Eq.~\eqref{eq:elastic_balance1} represents the torque balance 
for the bottom cylinder, while Eqs.~\eqref{eq:elastic_balance2} 
and \eqref{eq:elastic_balance3} correspond to the $x$-force balance 
for the bottom cylinder and the $y$-force balance for the upper 
cylinder, respectively.
Here, $\bm{n} = (n_x, n_y)$ and $\bm{t} = (t_x, t_y) = (n_y, -n_x)$ 
denote the unit normal and tangential vectors at the contact between 
the cylinders, where the subscripts $x$ and $y$ indicate the horizontal 
and vertical components, respectively, as shown in Fig.~\ref{fig1}(b).

We first consider the rigid-body case in which the cylinders do not 
deform. In this case, the relative positions of the cylinders remain 
fixed, and the unit normal and tangential vectors are 
$\bm{n} = (-1/2, \sqrt{3}/2)$ and $\bm{t} = (\sqrt{3}/2, 1/2)$, 
respectively. Solving 
Eqs.~(\ref{eq:elastic_balance1})--(\ref{eq:elastic_balance3}) 
together with the vertical force balance gives $d = (f + 3mg)/2$ and 
$e = \mu_c (f + mg)/2$, where $\mu_c \equiv 2 - \sqrt{3}$ denotes the 
critical friction coefficient at which the transition occurs, 
as explained below.

Using the above expressions, we derive the yield force in the 
rigid-body case, denoted $f_0$, as a function of the floor friction 
coefficient $\mu$, under the constraint $f\ge 0$. 
For $\mu < \mu_c /3$, the inequality $e > \mu d$ always holds, 
implying that slip occurs even at $f = 0$; thus, $f_0 = 0$. 
For $\mu_c /3 \leq \mu < \mu_c$, slip occurs at a finite 
$f_0$, where $e = \mu d$. For $\mu \ge \mu_c$, the inequality 
$e < \mu d$ always holds, and slip never occurs; hence, $f_0 = \infty$.
The mathematical expression for the yield force $f_0(\mu)$ is thus 
given by
\begin{align}\label{eq:yield_rigid_eng}
    f_0(\mu) =
    \begin{cases}
    0 & \mbox{for}\quad \mu < \dfrac{\mu_c}{3},\\[4pt]
    \dfrac{3\mu - \mu_c}{\mu_c-\mu} mg & \mbox{for}\quad \dfrac{\mu_c}{3} \leq \mu < \mu_c,\\[4pt]
    \infty & \mbox{for}\quad \mu_c \leq  \mu,
    \end{cases}
\end{align}
where $\mu_c = 2 - \sqrt{3} \simeq 0.268$ is the critical friction 
coefficient at which the yield force diverges, above which  
slip no longer occurs regardless of the applied force.
In what follows, we focus on the regime $\mu \ge \mu_c/3$.

The above analysis assumes ideal rigid bodies that do not deform 
under load. In contrast, real materials 
can be elastic and deform under 
load. This motivates us to investigate how the yield-force transition 
emerges in elastic systems, how their behavior asymptotically 
approaches the rigid-body limit, whether this limit truly 
coincides with the rigid-body case, and whether the magnitude 
of the yield force can be quantitatively predicted from material 
parameters. To address these questions, we analyze the system 
using an elastic model based on the discrete element method 
(DEM), both numerically and theoretically.

\sectionprl{Numerical results}
We investigate how the yield-force transition observed in the 
rigid-body model manifests in elastic systems. In this Letter, 
we employ a standard framework for frictional elastic bodies, 
based on DEM \cite{Cundall1979,Otsuki2011}. 
Specifically, we adopt a two-dimensional model with linear 
elastic forces and dissipation.
In this linear model, the contact forces are proportional to 
the normal and tangential displacements at the contact point, 
and the torque arm is approximated by the undeformed cylinder 
radius $r$. In principle, elastic deformation may modify the 
effective torque arm as $R = r + r_1\delta_n$, where $\delta_n$ 
denotes the normal displacement at contact and $r_1$ is the 
coefficient of the correction associated with this normal 
displacement. The correction term $r_1\delta_n$ would introduce 
a contribution to the torque that is quadratic in the displacements.
In the present analysis, this quadratic contribution is 
neglected by setting $r_1=0$.

Let us consider three cylinders, each having an identical mass $m$ and 
radius $r$, under gravity $g$. The position, velocity, and angular 
velocity of a cylinder $i$ are denoted by 
$\bm{r}_i$, $\bm{v}_i$, and $\omega_i$, respectively. 
Cylinder $i$ interacts with another cylinder $j$ when they overlap, 
i.e., $\Delta_{ij}\equiv 2r-r_{ij}>0$, where 
$\bm{r}_{ij}\equiv \bm{r}_i-\bm{r}_j=(x_{ij},y_{ij})$ and 
$r_{ij}=|\bm{r}_{ij}|$.
The contact force $\bm{f}_{ij}$ consists of normal and tangential 
components, $\bm{f}_{ij}^{(n)}$ and $\bm{f}_{ij}^{(t)}$, such that 
$\bm{f}_{ij}=\bm{f}_{ij}^{(n)} + \bm{f}_{ij}^{(t)}$. 
The normal contact force $\bm{f}_{ij}^{(n)}$ between 
cylinder $i$ and cylinder $j$ is given by 
$\bm{f}_{ij}^{(n)} = h_{ij}^{(n)} \, \Theta(\Delta_{ij}) \bm{n}_{ij}$, 
where $h_{ij}^{(n)}$ and $\bm{n}_{ij}$ are defined as 
$h_{ij}^{(n)} = k_n \Delta_{ij} - \eta_n v_{ij}^{(n)}$ and 
$\bm{n}_{ij} = \bm{r}_{ij} / |\bm{r}_{ij}|$, respectively. 
Here, $k_n$ and $\eta_n$ are the normal elastic and viscous constants,
and $v_{ij}^{(n)} \equiv (\bm{v}_i - \bm{v}_j) \cdot \bm{n}_{ij}$.
The Heaviside step function $\Theta(x)$ is defined as 
$\Theta(x)=1$ for $x\ge 0$ and $\Theta(x)=0$ otherwise. 
Similarly, the tangential contact force $\bm{f}_{ij}^{(t)}$ between 
cylinders $i$ and $j$ is given by 
$\bm{f}_{ij}^{(t)} = \min(\bm{h}_{ij}^{(t)}, -\mu |\bm{f}_{ij}^{(n)}| \bm{v}_{ij}^{(t)} / |\bm{v}_{ij}^{(t)}|) \Theta(\Delta_{ij})$, 
where $\min(\bm{a}, \bm{b})$ selects the vector with the smaller norm 
between $\bm{a}$ and $\bm{b}$, and 
$\bm{h}_{ij}^{(t)}=-k_t\bm{u}_{ij}^{(t)}-\eta_t \bm{v}_{ij}^{(t)}$.
Here, $k_t$ and $\eta_t$ are the tangential elastic and viscous 
constants. The tangential velocity $\bm{v}_{ij}^{(t)}$ and 
tangential displacement $\bm{u}_{ij}^{(t)}$ are respectively 
given by 
$\bm{v}_{ij}^{(t)} \equiv \left((\bm{v}_i - \bm{v}_j) \cdot \bm{t}_{ij} - r (\omega_i + \omega_j)\right) \bm{t}_{ij}$ 
where the tangential unit vector is defined as 
$\bm{t}_{ij} \equiv (-y_{ij} / r_{ij}, x_{ij} / r_{ij})$, 
and $\bm{u}_{ij}^{(t)}\equiv \int_{\mathrm{stick}}\bm{v}_{ij}^{(t)}d\tau$, 
where ``stick'' indicates that the integral is performed 
while $|\bm{h}_{ij}^{(t)}|<\mu|\bm{f}_{ij}^{(n)}|$.

\begin{figure*}[htbp]
    \centering
    \includegraphics[width=15cm]{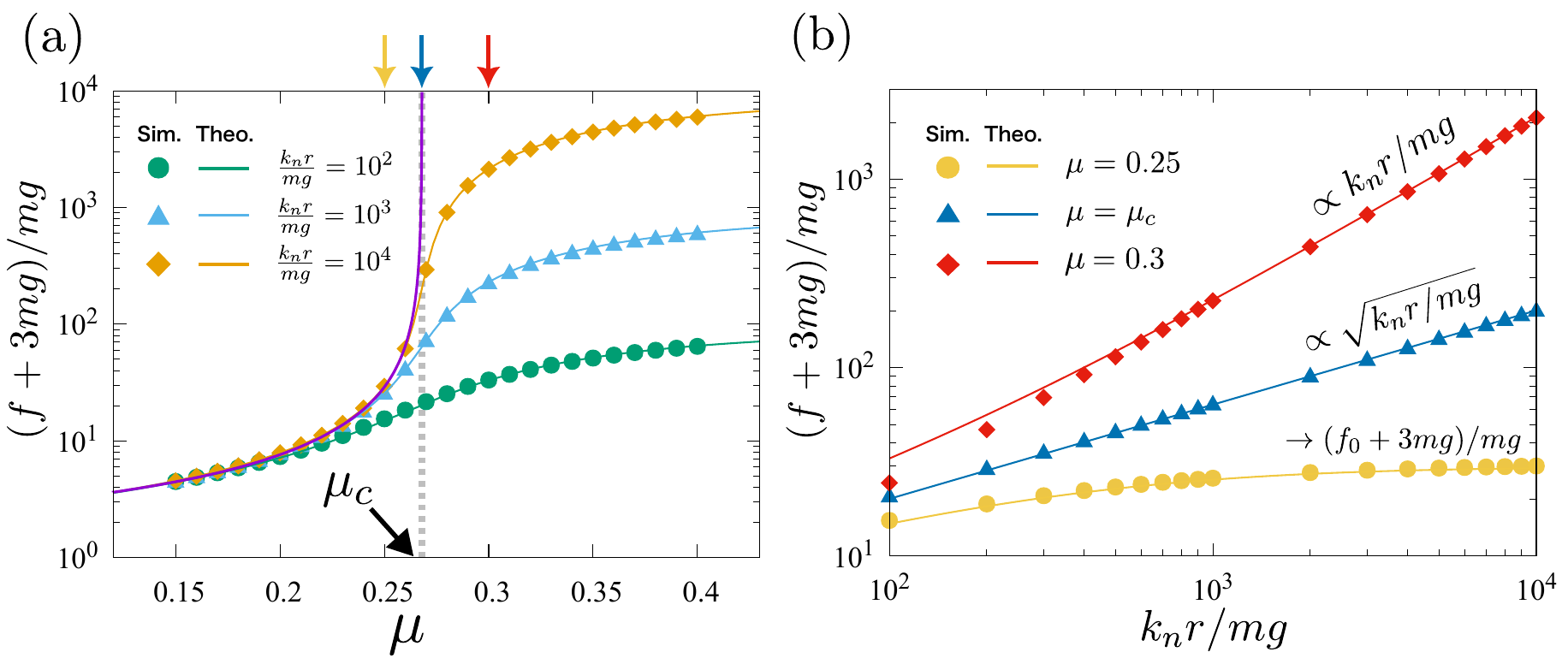}
    \caption{
        (a) Dependence of the yield force on the friction coefficient $\mu$. 
        Comparison among the rigid-body result (purple curve), 
        DEM simulations (symbols), and perturbative solutions (solid curves) 
        for $k_n r / mg= 10^2$, $10^3$, and $10^4$.
        (b) Scaling of the yield force with respect to the dimensionless stiffness $k_n r / mg$ 
        across the transition point, shown for $\mu = 0.25$, $\mu_c$, and $0.3$. Data symbols 
        represent DEM simulation results, and solid curves indicate 
        perturbative solutions. 
        Both panels show data for $\kappa=k_n/k_t=2.5$.
    }
    \label{fig2}
\end{figure*}

In numerical simulations, we measure the yield force as follows. 
We first prepare a stacked configuration of cylinders and then 
compress the system quasi-statically from above using a wall. 
The force $f(t)$ exerted on the wall is recorded during compression. 
Typically, the vertical reaction force increases with time, 
reaches a maximum, and then decreases. 
We define this peak value as the yield force $f(\mu, k_n r / mg)$, 
which depends on the friction coefficient $\mu$ and the dimensionless 
stiffness $k_n r / mg$. Details of the simulation parameters and 
the stacking and compression protocols are provided in \cite{supplement}.

Figure~\ref{fig2}(a) presents the measured yield force as a 
function of $\mu$ for $k_n r / mg = 10^2$, $10^3$, and $10^4$. 
While the sharp transition observed in the rigid model is 
smoothed out in the elastic case, an anomalous increase in 
the yield force is found near $\mu_c$. 
Figure~\ref{fig2}(b) shows the scaling of the yield force with 
respect to the dimensionless stiffness. 
The results show that the scaling of the yield force 
with respect to $k_n r / mg$ in the limit $k_n r / mg \gg 1$ differs
across the critical value $\mu_c$:
\begin{align}\label{eq:scaling}
    (f + 3mg)/mg = 
    \begin{cases}
        (f_0 + 3mg)/mg + O\!\left(mg/k_n r\right) & \mbox{for}\quad \mu<\mu_c,\\[4pt]
        O\!\left(\sqrt{k_n r/mg}\right) & \mbox{for}\quad \mu=\mu_c,\\[4pt]
        O\!\left(k_n r/mg\right) & \mbox{for}\quad \mu>\mu_c.
    \end{cases}
\end{align}
In the following, we show that a perturbative analysis accounts for 
both the anomalous enhancement and the scaling behavior observed 
in the simulations.

\sectionprl{Theoretical results}
We investigate the anomalous increase and the scaling of the yield 
force observed in numerical simulations by performing 
a perturbative analysis in the limit of small deformation. 
For this purpose, we introduce a small dimensionless parameter 
\begin{align}\label{eq:epsilon}
    \epsilon \equiv \frac{mg}{k_n r},
\end{align}
where $\epsilon = 0$ corresponds to the rigid-body limit. 
We fix the stiffness ratio $\kappa \equiv k_n / k_t$ and focus on the 
regime $\epsilon \ll 1$.

As shown in Fig.~\ref{fig3}, when an external force $f$ is applied 
from above, the bottom cylinders deform elastically and undergo a 
vertical displacement $z$. We now derive the equations and obtain 
a perturbative expression for the yield force.

\begin{figure}[htbp]
    \centering
    \includegraphics[width=4cm]{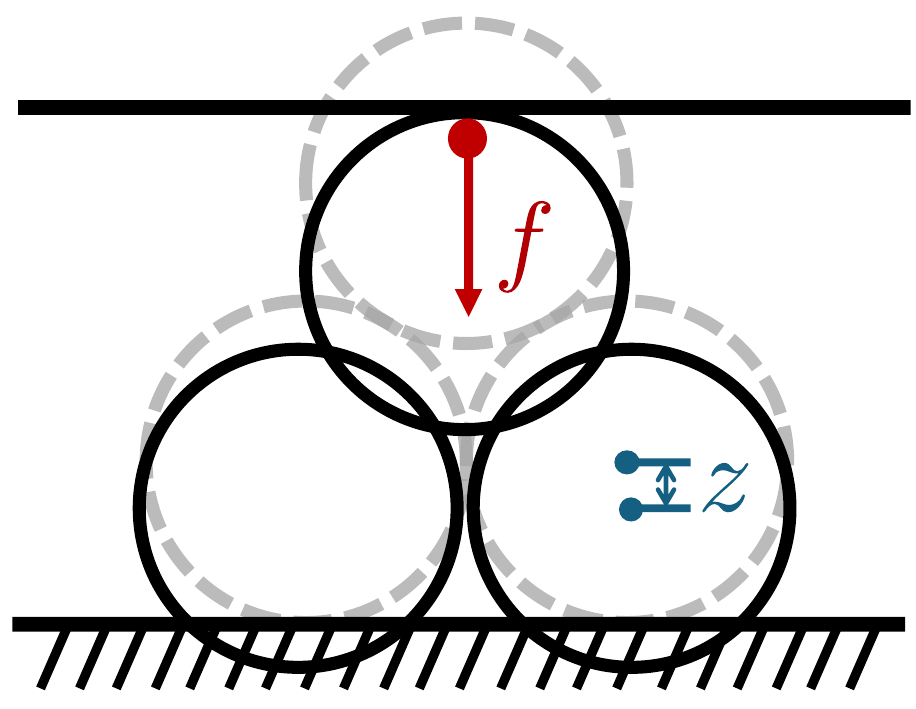}
    \caption{Displacement of the bottom cylinders under an applied 
    force $f$. The bottom cylinders move vertically by a displacement $z$.}
    \label{fig3}
\end{figure}

First, the global vertical force balance gives 
\begin{align}
    \frac{f + 3mg}{mg}  = \frac{2}{\epsilon r} z. \label{eq:f}
\end{align}
Since $f \ge 0$, this implies $z \ge 3\epsilon r / 2$ . 
From Eq.~(\ref{eq:f}), the yield force $f$ can be obtained once $z$ is 
determined. Second, $z$ is determined from the following equation,  
derived from Eqs.~(\ref{eq:elastic_balance1})--(\ref{eq:elastic_balance3}):
\begin{align}\label{eq:eq_z}
    z = r\, \mathcal{D}(\mu,\alpha)\, \mathcal{E}(\mu,\alpha),
\end{align}
where we introduce a dimensionless parameter 
$\alpha \equiv \epsilon r / z$. See End Matter for the derivation of 
Eq.~(\ref{eq:eq_z}). The functions 
$\mathcal{D}(\mu,\alpha)$ and $\mathcal{E}(\mu,\alpha)$ are defined as 
\begin{align}
    \mathcal{D}(\mu,\alpha) &\equiv \frac{\mu^2 + 2\sqrt{3}(1-\alpha)\mu -(1-\alpha)^2}{\mu\left((1-\alpha)^2+\mu^2\right)},\label{eq:D}\\[4pt]
    \mathcal{E}(\mu,\alpha) &\equiv \frac{(1-\alpha)^2}{(1-\alpha)^2+\kappa\left((1-\alpha)^2+\mu^2\right)}.\label{eq:E}
\end{align}
The yield force $f$ is obtained by solving Eq.~(\ref{eq:eq_z}) for $z$ 
under the constraint $z\ge 3\epsilon r/2$ and 
substituting the result into the force balance equation, Eq.~(\ref{eq:f}). 
Thus, our task reduces to solving Eq.~(\ref{eq:eq_z}) with $z\ge 3\epsilon r / 2$.

We now analyze Eq.~(\ref{eq:eq_z}) 
in the limit $\epsilon \to 0$. This equation can be 
rewritten as the following quintic equation: 
\begin{equation}
    \begin{aligned}\label{eq:poly}
    z^5 - (z_\star + \epsilon c_4^{(1)}) z^4
    + c_3 \epsilon z^3
    + c_2 \epsilon^2 z^2
    + c_1 \epsilon^3 z
    + \epsilon^4 = 0,
\end{aligned}
\end{equation}
where the coefficients are given by 
$c_i \equiv c_i^{(0)} + \epsilon c_i^{(1)}$, with $c_i^{(j)}$ 
being functions of $\mu$. Here, $z_\star$ is given by 
\begin{align}\label{eq:z_star}
    z_\star 
    = r\, \frac{(\mu - \mu_c)(\mu + \mu_c^{-1})}{\mu(1 + \mu^2)}\, \frac{1}{1 + \kappa(1 + \mu^2)}.
\end{align}
The function $z_\star$ changes sign across the transition point $\mu_c$, 
which leads to qualitatively different behavior of the solutions 
to Eq.~(\ref{eq:poly}).

We first consider a naive perturbative expansion and assume that the 
solution to Eq.~(\ref{eq:poly}) takes the form 
$z=C_0 \epsilon^\beta+o(\epsilon^\beta)$, where $C_0 \neq 0$ and 
$\beta \ge 0$ denote the leading-order coefficient 
and exponent of $z$ in the expansion with respect to $\epsilon$. 
Whether this ansatz is appropriate is confirmed a 
posteriori by checking that it yields a consistent solution.
By examining the order of each term in Eq.~(\ref{eq:poly}) with 
respect to $\epsilon$, the exponent $\beta$ is determined from the 
condition that at least two leading terms balance---a procedure known 
as the dominant balance method \cite{Bender2013}. 
Since the leading-order term changes depending on whether $z_\star$ 
vanishes (i.e., $z_\star=0$ at $\mu=\mu_c$ and $z_\star\neq 0$ otherwise), 
we separately consider the cases $\mu=\mu_c$ and $\mu\neq \mu_c$.
For $\mu=\mu_c$, the balances occur at $\beta = 1/2$ and $\beta = 1$. 
For $\beta = 1/2$, the terms $z^5$ and $c_3^{(1)} \epsilon z^3$ balance, 
yielding the solution 
$z = \pm r \sqrt{\epsilon / \mu_c(1 + 4\kappa \mu_c)} + o(\epsilon^{1/2})$.
For $\beta = 1$, we obtain $z = \epsilon r + o(\epsilon)$ and 
$z = \epsilon r / (4\mu_c) + o(\epsilon)$. Among these, the only solution 
satisfying $z \ge 3\epsilon r / 2$ is $z = r \sqrt{\epsilon / \mu_c(1 + 4\kappa \mu_c)} + o(\epsilon^{1/2})$.
For $\mu \neq \mu_c$, the same procedure gives the solutions consistent 
with $z \ge 3\epsilon r / 2$. Collecting the results, the naive 
perturbative solution of Eq.~(\ref{eq:eq_z}) yields 
\begin{align}\label{eq:naive_perturbation}
    z =
    \begin{cases}
        \mu_c \epsilon r/(\mu_c-\mu) + o(\epsilon), & \mu < \mu_c, \\
        r\sqrt{\epsilon/\mu_c(1+4\kappa\mu_c)} + o(\epsilon^{1/2}), & \mu = \mu_c, \\
        z_\star + o(1), & \mu > \mu_c,
    \end{cases}
    \quad \epsilon \to 0.
\end{align}
The scaling of the naive perturbative solution 
[Eq.~(\ref{eq:naive_perturbation})], combined with Eq.~(\ref{eq:f}), is 
consistent with the scaling relation Eq.~(\ref{eq:scaling}). 
This scaling further clarifies the physical meaning of $z_\star$: 
it represents the displacement at which 
slip occurs in the rigid body limit ($\epsilon \to 0$) for 
$\mu > \mu_c$. Since $z_\star$ remains $O(1)$ in this limit, 
the yield force diverges as the dimensionless stiffness 
$1/\epsilon$.

However, the solution diverges at the transition point, indicating the 
breakdown of the naive expansion near $\mu_c$. Therefore, 
an alternative expansion is required to correctly capture the 
singular behavior.

Now, we reformulate the perturbation theory by expanding 
Eq.~(\ref{eq:eq_z}) itself rather than its solution. 
Specifically, we assume the following expansion form 
of Eq.~(\ref{eq:eq_z}):
\begin{align}\label{eq:quadratic_z}
    z\left(z-z_\star\right) = \epsilon r^2 \Phi(\mu) + O(\epsilon^2).
\end{align}
The coefficient $\Phi(\mu)$ is obtained as
\begin{align}\label{eq:Phi}
    \Phi(\mu) =
    \begin{cases}
        F(\mu, 1-\mu/\mu_c) & \mbox{for}\quad \mu < \mu_c, \\
        F(\mu, 0) & \mbox{for}\quad \mu \ge \mu_c,
    \end{cases}
\end{align}
with
\begin{align}\label{eq:F}
    F(\mu,\alpha)
    &= \mathcal{E}(\mu,0)D(\mu,\alpha) \notag\\
    &\quad + \mathcal{D}(\mu,0)E(\mu,\alpha)+\alpha D(\mu,\alpha)E(\mu,\alpha),
\end{align}
where we introduce the functions $D(\mu,\alpha)$ and $E(\mu,\alpha)$ by
$\mathcal{D}(\mu,\alpha) = \mathcal{D}(\mu,0) + \alpha D(\mu,\alpha)$ and 
$\mathcal{E}(\mu,\alpha) = \mathcal{E}(\mu,0) + \alpha E(\mu,\alpha)$. 
See End Matter for the derivation of Eq.~(\ref{eq:quadratic_z}) together with 
Eqs.~(\ref{eq:Phi}) and (\ref{eq:F}).

Using Eqs.~(\ref{eq:quadratic_z}) and (\ref{eq:Phi}), we obtain 
the perturbative solution for the yield force $f_\epsilon$ as 
\begin{align}\label{eq:perturbative_solution}
    \frac{f_\epsilon + 3mg}{mg} = \frac{1}{\epsilon r} \left(z_\star + \sqrt{z_\star^2 + 4\epsilon r^2 \Phi(\mu)}\right).
\end{align}
In the rigid-body limit $\epsilon \to 0$, this expression reproduces 
the same scaling behavior as Eq.~(\ref{eq:scaling}).

Figure~\ref{fig2} compares the yield force $f_0$ obtained from 
the rigid-body case, the DEM simulation results $f(\mu, k_n r / mg)$, 
and the perturbative solution $f_\epsilon$. The DEM results 
agree well with the singular perturbative solution. 
Both $f(\mu, k_n r / mg)$ and $f_\epsilon$ 
approach the rigid-body result $f_0(\mu)$ in the limit 
$k_n r / mg \to \infty$. 
Moreover, the scaling behavior of the DEM simulation data with 
respect to the stiffness parameter $k_n r / mg$ ($=\epsilon^{-1}$) 
matches the singular perturbation scaling given 
in Eq.~(\ref{eq:scaling}).

\sectionprl{Concluding remarks}
In this Letter, we investigated the mechanical failure of a 
friction-stabilized system consisting of three cylinders stacked 
under gravity. In the rigid-body case, we identified a 
critical friction coefficient $\mu_c = 2 - \sqrt{3}$ that separates 
destructive and non-destructive regimes. 
We then examined the behavior in more realistic settings by performing 
DEM simulations for elastic cylinders and found that the yield 
force exhibits a singular increase near $\mu_c$. This anomalous 
behavior was analyzed using a singular perturbation approach, 
which yielded an explicit expression for the yield force, 
Eq.~(\ref{eq:perturbative_solution}). 
This expression quantitatively predicts the dependence on 
the DEM stiffness parameters and successfully 
captures the scaling behavior observed in the simulations.

We also remark on the scaling behavior described by 
Eq.~(\ref{eq:scaling}). Whether the associated exponent is 
an artifact of the linear-spring contact model used in our DEM 
or a universal value independent of the specific force law remains 
an open question. 
Exploring the robustness of this exponent across alternative 
contact models would be a natural direction for future work.
In addition, as detailed in End Matter, our singular perturbation 
analysis breaks down in the high-friction regime ($\mu \gtrsim 0.58$); 
developing a complementary treatment for that regime is left for 
future work.

We also address the experimental feasibility of observing 
this anomalous enhancement. The governing parameter 
$\epsilon=mg/(k_n r)$ can be estimated as $\epsilon \sim L\rho g/E$ 
from linear elasticity, where $E$, $\rho$, and $L$ denote 
the Young's modulus, mass density, and a representative length scale, 
respectively. For typical elastomeric cylinders 
($E\sim 1\,\mathrm{MPa}$, $\rho\sim10^3\,\mathrm{kg/m^3}$, 
$L\sim1\,\mathrm{cm}$), one obtains $\epsilon\sim10^{-4}$, 
which falls within the range studied here. The corresponding yield 
force is $f\simeq\epsilon^{-1}mg\sim10^2\,\mathrm{N}$, 
which is experimentally accessible. In contrast, for a wooden 
cylinder of the same mass ($m=1\,\mathrm{g}$), 
the much smaller value of $\epsilon$ leads to a yield force of order 
$f\simeq\epsilon^{-1}mg\simeq10^5\,\mathrm{N}$ for $\mu>\mu_c$. 
Therefore, the anomalous enhancement above $\mu_c$ can be 
realistically observed using sufficiently soft materials, and 
elastomers represent a viable platform for this purpose, provided 
that their surface friction is properly controlled. 
However, such experimental control is not straightforward and is 
beyond the scope of the present paper; it will be addressed in a 
future publication.

Finally, we remark on the relation between the instability 
identified here and the mechanical behavior of larger 
granular assemblies. The failure studied in this work 
corresponds to an instability of a minimal structural 
unit: the onset of sliding at the floor contacts of a 
three-body stack. This is distinct from the macroscopic 
failure or plastic deformation of a bulk granular material, 
which involves contact-network rearrangements.
Therefore, a direct correspondence between our results 
and bulk material behavior should not be assumed without 
further analysis. Nevertheless, we argue that the three-body 
system studied here plays a role analogous to that of 
few-body problems in other areas of physics. 
In the kinetic theory of gases, 
the chaotic instability of three-body trajectories 
provided the dynamical foundation for the molecular chaos 
hypothesis underlying macroscopic transport phenomena. 
By analogy, the critical transition identified here in 
the three-body problem---the existence of a critical 
friction coefficient $\mu_c$ and the associated scaling 
law---may capture a fundamental aspect of the mechanical 
response of more general granular assemblies. 
How this transition manifests in, or is 
modified by, many-body interactions remains an important 
open problem, which we identify as a central direction 
for future work.

\smallskip
\sectionprl{Acknowledgements}
The authors thank K.~Yokota, Y.~Yanagisawa, A.~Yoshida, 
M.~Itami, S.~Inagaki, M.~Otsuki, S.~Poincloux, and 
D.~S.~Shimamoto for fruitful discussions.
This study was supported by JSPS KAKENHI Grant Numbers 
JP22K13975, JP23K22415, JP25K00923, JP26H00383 
and JST SPRING Grant Number JPMJSP2110.

%% End Matter
\newpage
\appendix

\onecolumngrid
\vspace{5mm}
\begin{center}
    \large \textbf{End Matter}
\end{center}
\twocolumngrid

\sectionprl{Appendix A: Derivation of Eq.~(\ref{eq:eq_z})}
\renewcommand{\theequation}{E\arabic{equation}}
\setcounter{equation}{0}
We first describe the elastic deformation by four degrees 
of freedom, as illustrated in Fig.~\ref{figEM}: 
the vertical displacements $\delta_{11}$ and $\delta_{21}$ of 
cylinders 1 and 2, the horizontal displacement $\delta_{22}$ of 
cylinder 2, and its rotation angle $\theta$.
In elastic systems, deformations alter the relative positions between 
cylinders, allowing slip to occur even for $\mu>\mu_c$.
Within the DEM framework, the contact normal and tangential 
vectors $\bm{n}$ and $\bm{t}$ are expressed in terms of the displacement 
variables $\delta_{11}$, $\delta_{21}$, $\delta_{22}$, and $\theta$. 
Similarly, the corresponding contact forces can also be expressed as functions of 
these displacements. Substituting these expressions into the mechanical equilibrium 
conditions and the friction criterion, 
Eqs.~(\ref{eq:elastic_balance1})--(\ref{eq:elastic_balance3}), 
we obtain five equations for the five unknowns: 
$\delta_{11}$, $\delta_{21}$, $\delta_{22}$, $\theta$, and the yield 
force $f$. The explicit forms of the forces and unit vectors are 
provided in \cite{supplement}.

\begin{figure}[htbp]
    \centering
    \includegraphics[width=5cm]{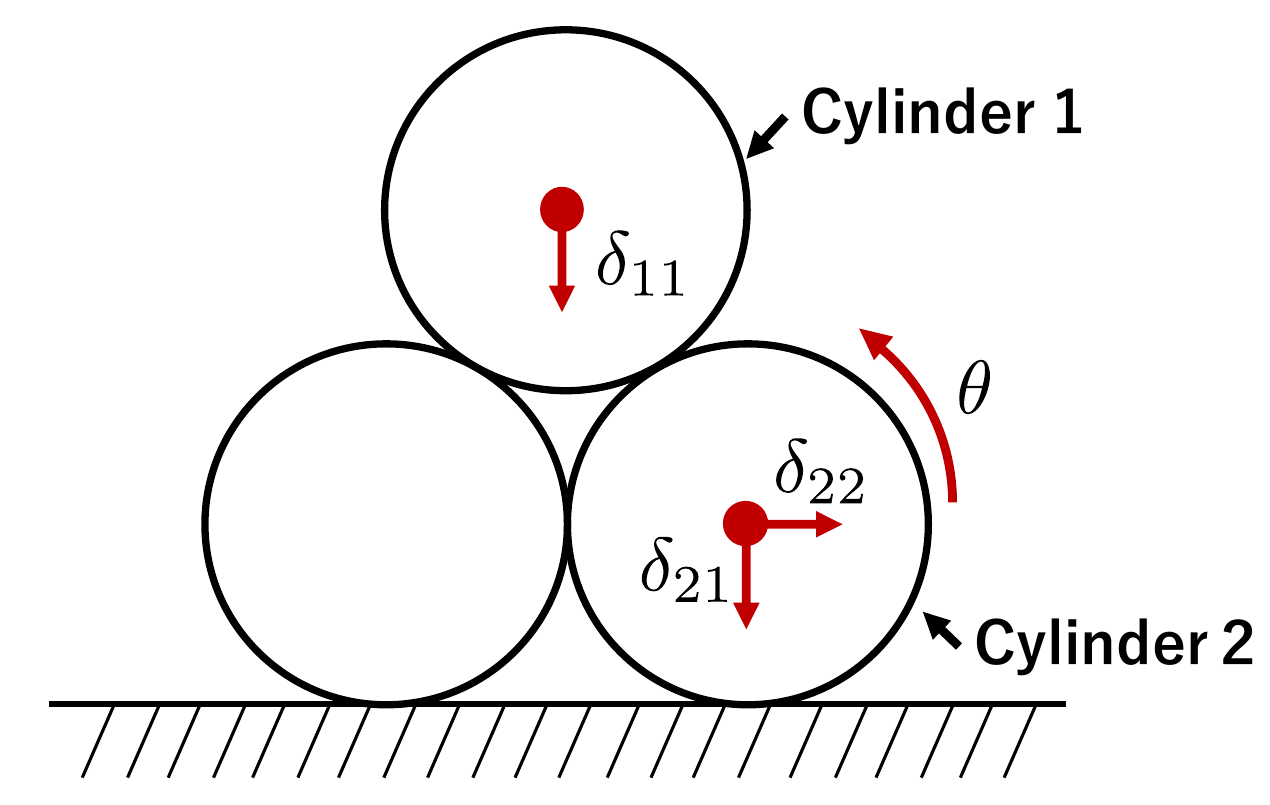}
    \caption{Displacement variables under an applied force $f$. 
    Cylinder~1 undergoes vertical displacement $\delta_{11}$, 
    while cylinder~2 displaces vertically by $\delta_{21}$, horizontally by 
    $\delta_{22}$, and rotates by an angle $\theta$.}
    \label{figEM}
\end{figure}

We introduce new variables: 
$x\equiv \delta_{11} - \delta_{21}$, $y \equiv \delta_{22}$, and 
$z \equiv \delta_{21}$, where all variables are non-negative to 
ensure that the corresponding contact forces remain positive. 
The slip condition $e = \mu d$ 
leads to an additional constraint expressed as
\begin{align}\label{eq:theta}
    r \theta = \mu \kappa z - y,
\end{align}
which allows us to eliminate $\theta$ from Eqs.~(\ref{eq:elastic_balance1})--(\ref{eq:elastic_balance3}).
We then rewrite Eqs.~(\ref{eq:elastic_balance1})--(\ref{eq:elastic_balance3}) 
in terms of $x$, $y$, and $z$ as
\begin{align}
    &\sqrt{(\sqrt{3}r-x)^2 + (r+y)^2}(2\mu\kappa z-y) = r(x+\sqrt{3}y),\label{eq:xyz_1}\\
    &A(r+y) -B(\sqrt{3}r-x) -\mu C = 0,\label{eq:xyz_2}\\
    &A(\sqrt{3}r-x) + B(r+y)- C+ \epsilon r^3 (x+\sqrt{3}y)^2=0,\label{eq:xyz_3}
\end{align}
where the auxiliary functions $A$, $B$, and $C$ are defined as 
$A\equiv \left\{x(\sqrt{3}r-x) - y(r+y)\right\}(2\mu \kappa z-y)^2$, 
$B\equiv \mu r(x +\sqrt{3}y)(2\mu \kappa z-y)z$, and 
$C\equiv r^2(x+\sqrt{3}y)^2z$, respectively.
See Supplemental Material \cite{supplement} for the derivation of 
Eqs.~(\ref{eq:xyz_1})--(\ref{eq:xyz_3}).

Now, we attempt to eliminate $x$ and $y$ from 
Eqs.~(\ref{eq:xyz_1})--(\ref{eq:xyz_3}). First by combining 
Eqs.~(\ref{eq:xyz_2}) and (\ref{eq:xyz_3}) to eliminate the common 
term denoted by $A$, we obtain 
\begin{align}
    &\mu r \left\{(\sqrt{3}r-x)^2+(r+y)^2\right\}(x+\sqrt{3}y)(2\mu \kappa z-y)z \notag\\
    &+ r^2 \left\{(\sqrt{3}r-x)\mu -(r+y)\right\}(x+\sqrt{3}y)^2 z \notag\\
    &+ \epsilon r^3 (x+\sqrt{3}y)^2(r+y)=0,
\end{align}
where we have substituted the explicit forms of $B$ and $C$. 

Next, we use Eq.~(\ref{eq:xyz_1}) to eliminate the factor 
$(2\mu \kappa z -y)$ from the first term and divide both sides by 
$(x+\sqrt{3}y) \neq 0$, yielding
\begin{align}\label{eq:xyz_4}
    \mu z \left\{\sqrt{(\sqrt{3}r-x)^2 + (r+y)^2} + (\sqrt{3}r -x)\right\} = (r+y)(z-\epsilon r).
\end{align}
Multiplying both sides by the conjugate factor 
$\sqrt{(\sqrt{3}r-x)^2 + (r+y)^2} - (\sqrt{3}r -x)$ leads to 
the relation 
\begin{align}\label{eq:xyz_5}
    (z-\epsilon r) \left\{\sqrt{(\sqrt{3}r-x)^2 + (r+y)^2} - (\sqrt{3}r -x)\right\} = \mu (r+y)z.
\end{align}
Combining Eqs.~(\ref{eq:xyz_4}) and (\ref{eq:xyz_5}) allows us to 
eliminate the square roots, yielding 
\begin{align}\label{eq:x}
    r + y = \frac{(1-\alpha)^2 -\mu^2}{2\mu (1-\alpha)}(\sqrt{3}r-x), 
\end{align}
where we have used $\alpha = \epsilon r / z$, which can subsequently 
be used to eliminate $x$.

Substituting Eq.~(\ref{eq:x}) into Eq.~(\ref{eq:xyz_2}) to eliminate 
$x$, and canceling the nonzero factor $(r+y)$, we further eliminate 
$y$, ultimately yielding a closed-form expression for $z$: 
\begin{align}
    z = r\, \frac{\mu^2 + 2\sqrt{3}(1-\alpha)\mu -(1-\alpha)^2}{\mu((1-\alpha)^2 + \mu^2)} \frac{(1-\alpha)^2}{(1-\alpha)^2 + \kappa((1-\alpha)^2 + \mu^2)}.
\end{align}
This corresponds to Eq.~(\ref{eq:eq_z}) together with Eqs.~(\ref{eq:D}) and 
(\ref{eq:E}).

\bigskip
\sectionprl{Appendix B: Derivation of Eq.~(\ref{eq:quadratic_z})}
Starting from Eq.~(\ref{eq:eq_z}), we expand $z$ as
\begin{align}
    \begin{aligned}
        z
        &= r\mathcal{D}(\mu,\alpha)\mathcal{E}(\mu,\alpha)\\
        &= r\left(\mathcal{D}(\mu,0)+\alpha D(\mu,\alpha)\right)\left(\mathcal{E}(\mu,0)+\alpha E(\mu,\alpha)\right)\\
        &= z_\star +\alpha r (\mathcal{E}(\mu,0)D(\mu,\alpha) + \mathcal{D}(\mu,0)E(\mu,\alpha)\\
        &\qquad \qquad \qquad \qquad \qquad +\alpha D(\mu,\alpha)E(\mu,\alpha)), 
    \end{aligned}
\end{align}
where we have used $z_{\star}=r\,\mathcal{D}(\mu,0)\mathcal{E}(\mu,0)$. 
Subtracting $z_\star$ and substituting $\alpha = \epsilon r / z$, 
we obtain 
\begin{align}\label{eq:quadratic_z2}
    z(z - z_\star) = \epsilon r^2F(\mu,\alpha)
\end{align}
with Eq.~(\ref{eq:F}).
We then extract the leading-order contribution, 
$\Phi = \lim_{\epsilon\to 0} F(\mu,\alpha)$. Using the naive 
perturbative solution in Eq.~(\ref{eq:naive_perturbation}), we 
find 
\begin{align}
    \lim_{\epsilon\to 0} \alpha=
    \begin{cases}
        1-\mu/\mu_c, &\mu\leq \mu_c,\\
        0, & \mu>\mu_c.
    \end{cases}
\end{align}
Substituting this limit into $F(\mu,\alpha)$ yields 
Eq.~(\ref{eq:quadratic_z}) together with Eq.~(\ref{eq:Phi}).

\sectionprl{Appendix C: Limitations of the analysis}
We note here the limitations of our perturbative analysis for 
elastic case. In particular, the assumption that 
the contact force $c$ vanishes at the onset of 
failure---corresponding to $y > 0$---breaks down for 
sufficiently large friction coefficients. For $\mu \gtrsim 0.8$, 
the perturbative solution to Eqs.~(\ref{eq:xyz_1})--(\ref{eq:xyz_3}) 
within $y > 0$ ceases to exist, and a solution with $y < 0$ emerges, 
suggesting a regime where the structure remains 
non-destructive even in the elastic case. Furthermore, DEM 
simulations with $k_n r / mg = 10^2$ show that  for $\mu\gtrsim 0.58$, 
the upper wall comes into contact with cylinder 2 before sliding occurs 
at the base, making it impossible to define a yield force. 
These findings indicate that a different treatment is necessary 
in the high-friction regime, both theoretically and numerically.

% === Supplemental Material ===
\onecolumngrid

\clearpage
\begin{center}
\textbf{\large Supplemental Material for\\
``Anomalous Enhancement of Yield Strength due to Static Friction''}
\end{center}
\vspace{1em}

\renewcommand{\theequation}{S\arabic{equation}}
\setcounter{equation}{0}

\renewcommand{\thefigure}{S\arabic{figure}}
\setcounter{figure}{0}

\author{
Ryudo Suzuki$^{1}$,
Takashi Matsushima$^{2}$,
Tetsuo Yamaguchi$^{3}$,
Marie Tani$^{1}$,
Shin-ichi Sasa$^{1}$
}

\affiliation{$^1$Department of Physics, Kyoto University, Kyoto 606-8502, Japan}
\affiliation{$^2$Faculty of Engineering, Information and Systems, University of Tsukuba, Tsukuba, Ibaraki 305-8573, Japan}
\affiliation{$^3$Department of Biomaterial Sciences, The University of Tokyo, 1-1-1 Yayoi, Bunkyo-ku, Tokyo 113-8657, Japan}

\maketitle

This Supplemental Material provides detailed descriptions of the 
numerical model and the derivation of 
Eqs.~(\ref{eq:xyz_1})--(\ref{eq:xyz_3}) presented in the 
main text. In Section~I, we first present the complete 
implementation of the two-dimensional discrete element method (DEM), 
including the equations of motion, contact 
force laws, and frictional interactions. We then describe the 
protocol for preparing the initial stacked configuration and the 
procedure for quasi-static compression used to measure the yield force. 
In Section~II, we derive explicit expressions for 
contact forces and Eqs.~(\ref{eq:xyz_1})--(\ref{eq:xyz_3}) 
in the main text.
Finally, in Section~III, we provide explanations 
of Supplemental Movies~S1 and~S2, which show the compression 
experiments of stacked cylinders on different substrates.

\section{I. Details of DEM Simulation}
In this study, we employ a standard two-dimensional discrete element 
method (DEM). We first explain the parameter values used in 
the simulations, followed by the protocol for preparing 
the initial stacked configuration and the procedure for the 
quasi-static compression used to measure the yield force.

\subsection{A. Simulation Parameters}
We employ a leapfrog integration scheme. 
The spring constants are set at a 
fixed ratio of $k_n / k_t = 2.5$, and the normal stiffness is 
varied as $k_n r / mg = 10^2$, $10^3$, $10^4$. 
The inter-cylinder friction coefficient 
is fixed at $\mu_d = 0.7$, while the floor friction coefficient $\mu$ 
is treated as a control parameter. The damping coefficients 
$\eta_n$ and $\eta_t$ are determined to yield a restitution 
coefficient of $e_r=0.7$, given by 
\begin{align}
    \eta_n &= -2 \ln e_r \sqrt{\frac{mk_n}{\pi^2 +(\ln e_r)^2}}, \\
    \eta_t &= -2 \ln e_r \sqrt{\frac{mk_t}{\pi^2 +(\ln e_r)^2}}.
\end{align}

\subsection{B. Preparation of the Initial Stacked Configuration}
In the stack of three cylinders, the force balance conditions alone 
do not uniquely determine the contact forces; in particular, 
the contact force $c$ between the lower cylinders remains 
indeterminate. To specify the initial state, we 
prepare a mechanically equilibrated configuration with $c=0$ 
for each choice of stiffness and friction coefficients.

\begin{figure}[h]
    \centering
    \includegraphics[width=0.6\linewidth]{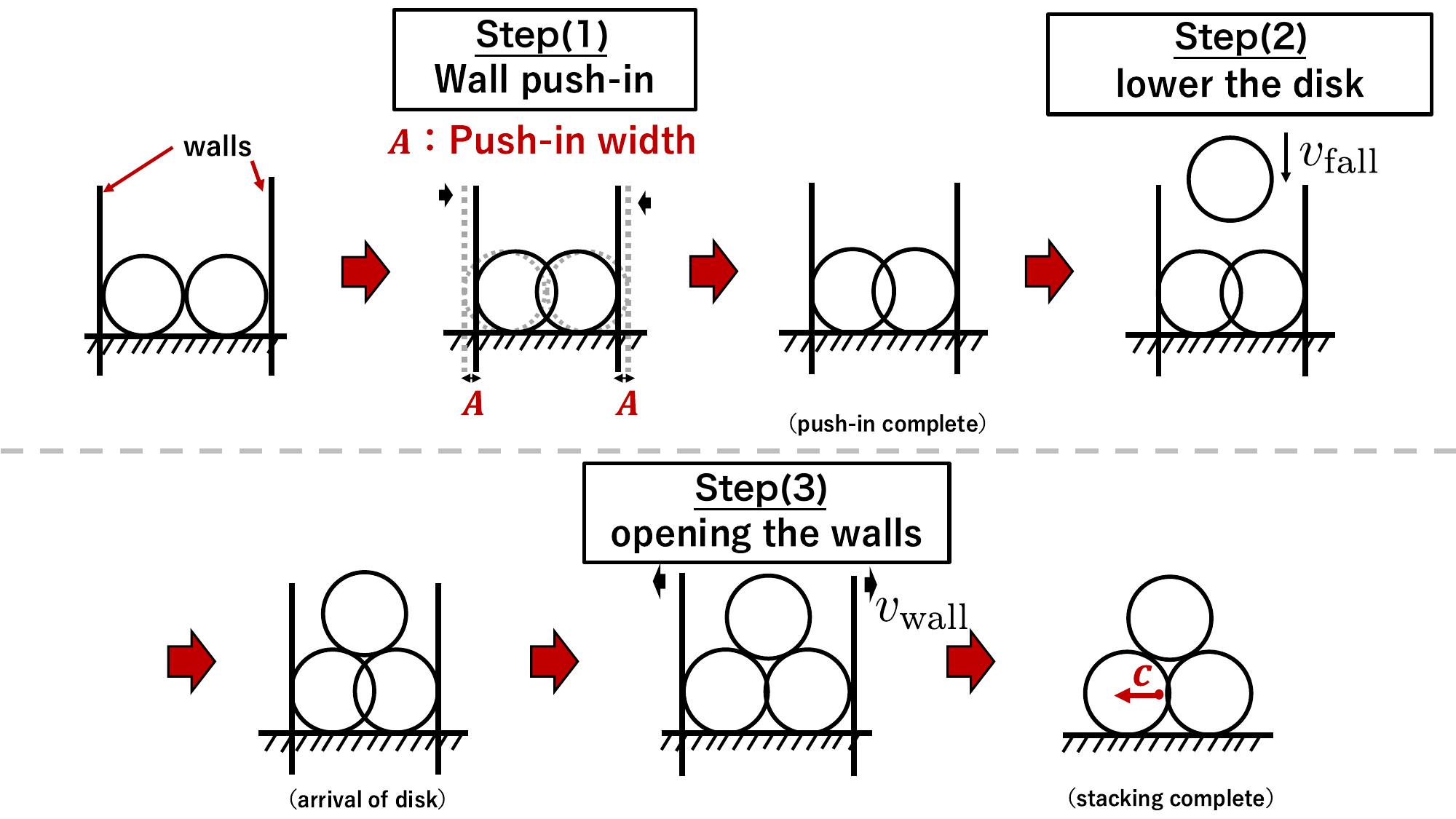}
    \caption{
    Preparation protocol for the initial state. The stacking 
    procedure consists of three steps: 
    (1) inward displacement of the walls by a distance $A$;
    (2) release of the top cylinder at a constant velocity 
    $v_{\mathrm{fall}}$; and 
    (3) retraction of the walls at a constant velocity 
    $v_{\mathrm{wall}}$.
    }
    \label{fig:protocol}
\end{figure}

The preparation protocol, illustrated in Fig.~\ref{fig:protocol}, 
consists of three steps and employs frictionless walls to suppress 
tangential motion during wall manipulation: (1) the walls are 
pushed inward by a distance $A$; (2) the top cylinder is released 
at a constant falling speed $v_{\mathrm{fall}}$; and (3) the walls 
are retracted at a constant speed $v_{\mathrm{wall}}$.

The falling speed of the top cylinder is fixed at 
$(v_{\mathrm{fall}}/ r) \sqrt{m/k_n}= 0.1$, and the retraction 
speed of the wall at $(v_{\mathrm{wall}}/ r) \sqrt{m/k_n} = 10^{-3}$. 
This protocol enables systematic preparation of static configurations 
characterized by different values of the internal contact variable 
$c$, by varying the control parameter $A$. In particular, we 
generate the special case of $c=0$ and evaluate the yield force.

\subsection{C. Quasi-Static Compression Protocol}
\begin{figure}[h]
    \centering
    \includegraphics[width=8cm]{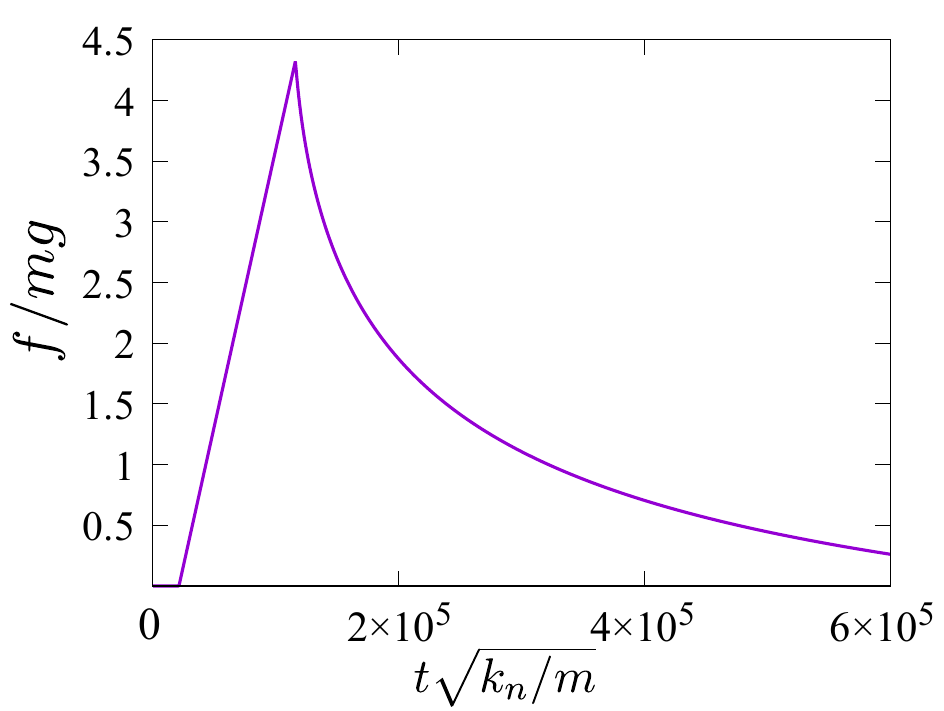}
    \caption{
        Time series of the compressive force $f(t)$ acting on the 
        top wall during compression. The peak value is defined 
        as the yield force $f(\mu, k_n r / mg)$.
        The parameters are $\mu = 0.2$, $k_n r/mg = 10^2$, 
        and $(v_{\mathrm{com}} / r)\sqrt{m / k_n} = 10^{-6}$.
        }
    \label{fig:yield_force}
\end{figure}

The yield force was evaluated by applying a quasi-static 
compression protocol to the stacked configuration prepared as 
described above. Specifically, the structure was compressed from 
above using a wall moving at a constant speed $v_{\mathrm{com}}$, 
and the yield force was defined as the peak value of the compressive 
force during this process, as shown in Fig.~\ref{fig:yield_force}. 

\begin{figure}[h]
    \centering
    \includegraphics[width=13cm]{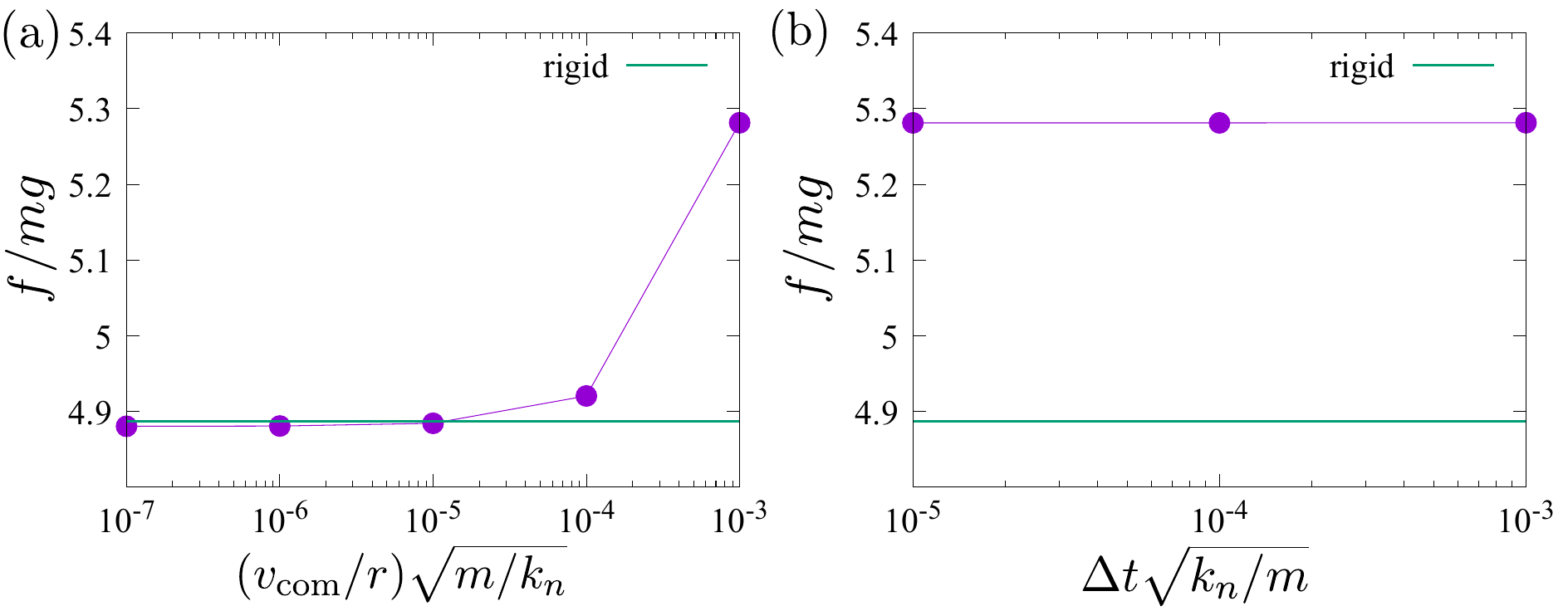}
    \caption{
        Dependence of the yield force on the compression speed 
        $v_{\mathrm{com}}$ and the time step $\Delta t$. 
        (a) Yield force versus $v_{\mathrm{com}}$ for fixed 
        friction coefficient $\mu = 0.2$, stiffness 
        $k_n r / mg = 10^4$, and time step $\Delta t \sqrt{k_n / m} = 10^{-3}$. 
        The green line indicates the rigid-body limit value $f_0$. 
        The yield force converges to a constant value at 
        $(v_{\mathrm{com}}/r) \sqrt{m / k_n} =10^{-6}$. 
        (b) Yield force versus $\Delta t$ for fixed 
        $\mu = 0.2$, $k_n r / mg = 10^4$, and 
        $(v_{\mathrm{com}}/r) \sqrt{m / k_n} = 10^{-3}$. 
        The yield force remains unchanged for 
        $\Delta t \sqrt{k_n / m} \leq 10^{-3}$, indicating that 
        $\Delta t \sqrt{k_n / m} = 10^{-3}$ is sufficient for the measurement.
        }
    \label{fig:vwalldt}
\end{figure}

Quasi-static conditions were ensured by choosing the compression 
speed to be sufficiently small. As shown in Fig.~\ref{fig:vwalldt}(a), 
the measured yield force converges as $v_{\mathrm{com}}$ decreases. 
Based on this convergence, we fixed the compression speed at 
$(v_{\mathrm{com}} / r) \sqrt{m / k_n} = 10^{-6}$. 
The time step was set to $\Delta t \sqrt{k_n / m} = 10^{-3}$, which is 
sufficiently small to ensure convergence, as confirmed in 
Fig.~\ref{fig:vwalldt}(b).

\section{II. Derivation of Eqs.~(\ref{eq:xyz_1})--(\ref{eq:xyz_3})}
In this section, we derive explicit expressions for the contact 
forces and for Eqs.~(\ref{eq:xyz_1})--(\ref{eq:xyz_3}) in the main text. 
To this end, we consider the small-displacements limit, in which 
the expression for tangential displacement in the DEM simplifies, 
as described below.

To define the contact forces, 
we first specify the displacement vectors of cylinders 1 and 2 
\begin{align}
    \bm{\delta}_1 = 
    \begin{pmatrix}
        0\\
        -\delta_{11}
    \end{pmatrix},
    \quad
    \bm{\delta}_2 =
    \begin{pmatrix}
        \delta_{22}\\
        -\delta_{21}
    \end{pmatrix}.
\end{align}
Using these, the post-deformation center-of-mass positions of 
cylinders 1 and 2 are given by 
\begin{align}
    \bm{r}_1 = 
    \begin{pmatrix}
        0\\
        (1+\sqrt{3})r
    \end{pmatrix}
    + \bm{\delta}_1 =
    \begin{pmatrix}
        0\\
        (1+\sqrt{3})r-\delta_{11}
    \end{pmatrix},
    \quad
    \bm{r}_2 = 
    \begin{pmatrix}
        r\\
        r
    \end{pmatrix}
    +\bm{\delta}_2
    = 
    \begin{pmatrix}
        r+\delta_{22}\\
        r-\delta_{21}
    \end{pmatrix}.
\end{align}
The relative position vector from cylinder 2 to cylinder 1 is then given by
\begin{align}
    \bm{r}_{21} = \bm{r}_1 -\bm{r}_2
    = \begin{pmatrix}
        -r-\delta_{22}\\
        \sqrt{3}r +\delta_{21} -\delta_{11}
    \end{pmatrix}.
\end{align}
The normal and tangential unit vectors at the contact are expressed as 
\begin{align}
    \bm{n} &= \frac{\bm{r}_{21}}{|\bm{r}_{21}|} 
    = \frac{1}{\sqrt{(r+\delta_{22})^2 +(\sqrt{3}r+\delta_{21}-\delta_{11})^2}}
    \begin{pmatrix}
        -r-\delta_{22}\\
        \sqrt{3}r + \delta_{21}-\delta_{11}
    \end{pmatrix},\\
    \bm{t} &=
    \begin{pmatrix}
        n_y\\
        -n_x
    \end{pmatrix}
    = \frac{1}{\sqrt{(r+\delta_{22})^2 +(\sqrt{3}r+\delta_{21}-\delta_{11})^2}}
    \begin{pmatrix}
        \sqrt{3}r + \delta_{21}-\delta_{11}\\
        r+\delta_{22}
    \end{pmatrix}.
\end{align}

To compute the contact force at the contact points, we first 
evaluate the relative displacement between the cylinders, considering 
only translational contributions. The relative displacement vector 
from cylinder 1 to 2 is given by 
\begin{align}
    \bm{\delta}_{12} = \bm{\delta}_2 -\bm{\delta}_1 = 
    \begin{pmatrix}
        \delta_{22}\\
        \delta_{11} -\delta_{21}
    \end{pmatrix}.
\end{align}
Including the rotational contribution, the relative displacement of 
cylinder 1 with respect to cylinder 2 at the contact point decomposes 
into normal and tangential components. The normal component is 
\begin{align}
    -(\bm{\delta}_{12}\cdot \bm{n})\bm{n},
\end{align}
while the tangential component reads 
\begin{align}
    (-(\bm{\delta}_{12}\cdot \bm{t})+ r\theta)\bm{t}.
\end{align}
In our model, the contact force is defined by applying linear springs 
with stiffness $k_n$ and $k_t$ in the normal and tangential 
directions, respectively. The resulting contact force components 
$a$ and $b$ are thus given by 
\begin{align}
    a &= k_n (\bm{\delta}_{12}\cdot \bm{n}),\label{eq_SM:a}\\
    b &= k_t[(\bm{\delta}_{12}\cdot \bm{t})-r\theta].\label{eq_SM:b}
\end{align}

We next compute the relative displacement at the contact point 
between cylinder 2 and the floor. Let $\bm{e}_x$ and $\bm{e}_y$ 
denote the unit vectors along the $x$- and $y$-axes, respectively. 
The tangential (horizontal) relative displacement is 
\begin{align}
    ((\bm{\delta}_2\cdot \bm{e}_x)+r\theta)\bm{e}_x,
\end{align}
while the normal (vertical) relative displacement is 
\begin{align}
    (\bm{\delta}_2\cdot \bm{e}_y) \bm{e}_y.
\end{align}
The corresponding contact forces $d$ and $e$ are then given by 
\begin{align}
    d &= -k_n(\bm{\delta}_2\cdot \bm{e}_y),\label{eq_SM:d}\\
    e &= k_t[(\bm{\delta}_2\cdot \bm{e}_x)+r\theta].\label{eq_SM:e}
\end{align}
Substituting the explicit forms of $\bm{n}$, $\bm{t}$, and 
$\bm{\delta}_2$, the contact force components $a$, $b$, $d$, and $e$ 
are expressed as 
\begin{gather}
    a = \frac{1}{\sqrt{(r+\delta_{22})^2 +(\sqrt{3}r+\delta_{21}-\delta_{11})^2}} k_n\left(-\delta_{22}(r+\delta_{22}) + (\delta_{11}-\delta_{21})(\sqrt{3}r+\delta_{21}-\delta_{11})\right),\\
    b = k_t \left(\frac{1}{\sqrt{(r+\delta_{22})^2 +(\sqrt{3}r+\delta_{21}-\delta_{11})^2}} (\sqrt{3}\delta_{22}+\delta_{11}-\delta_{21})-r\theta\right),\\
    d = k_n \delta_{21},\\
    e = k_t(\delta_{22}+r\theta).
\end{gather}

Substituting the explicit expressions of $a$, $b$, $d$, $e$, 
as well as the unit vectors $\bm{n}$ and $\bm{t}$ into the force 
balance equations, 
\begin{gather}
    b - e = 0,\label{eq_SM:eq3}\\
    a n_x + b(t_x + 1) = 0,\label{eq_SM:eq4}\\
    2 a n_y + 2b t_y -f -mg = 0.\label{eq_SM:eq5}
\end{gather}
and introducing the new variables $x=\delta_{11}-\delta_{21}, 
y=\delta_{22}$, and $z=\delta_{21}$, we obtain 
Eqs.~(\ref{eq:xyz_1})--(\ref{eq:xyz_3}) presented in End Matter.

\section{III. Supplemental Movies S1 and S2}
\begin{figure}[h]
    \centering
    \includegraphics[width=10cm]{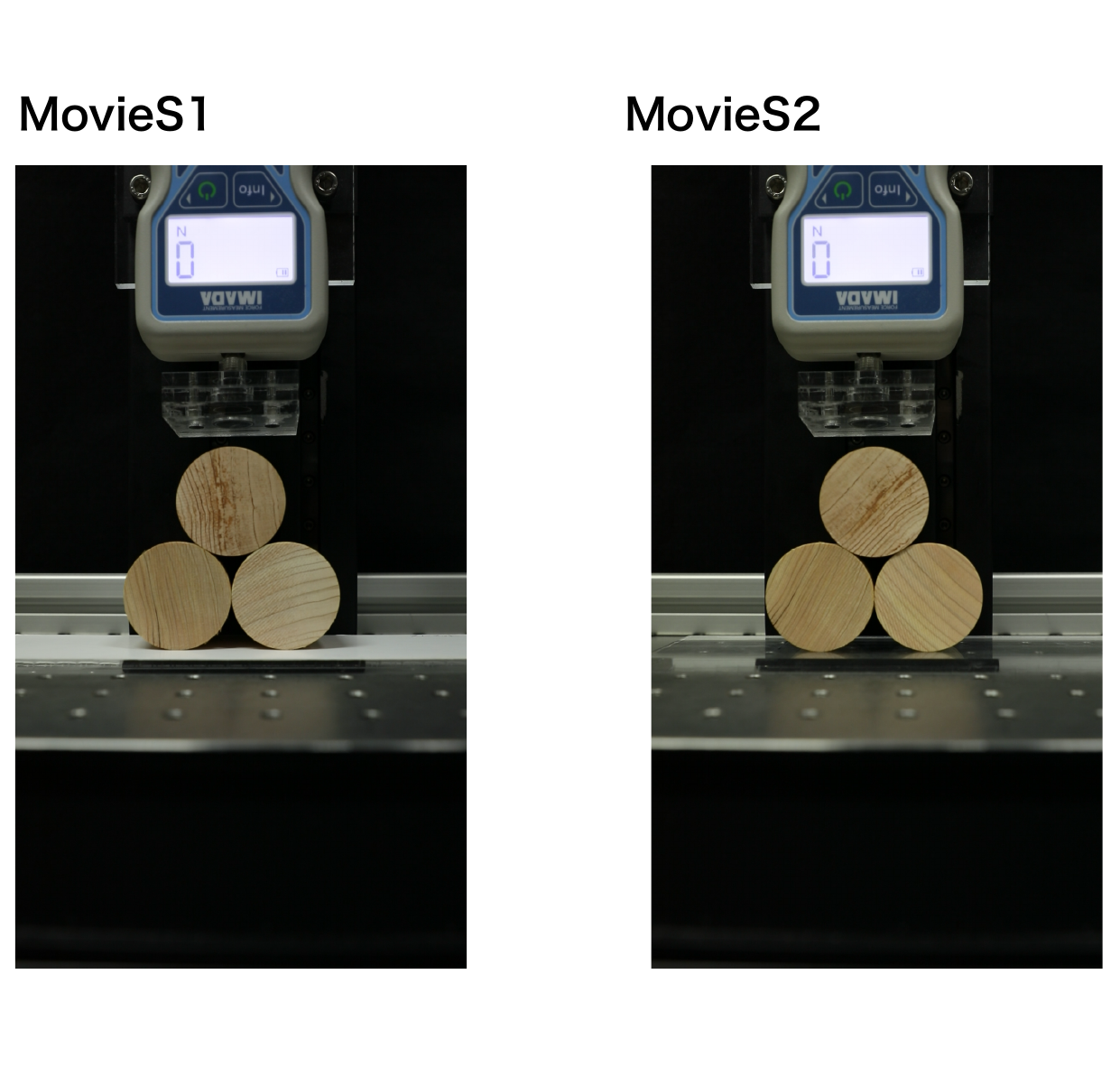}
    \caption{
    Movie S1: Compression experiment on snow paper.
    Movie S2: Compression experiment on an acrylic plate.
    }
    \label{fig:Movie}
\end{figure}

Supplemental Movies S1 and S2 show the compression experiments of 
stacked cylindrical particles. 
Each cylinder is made of wood with a radius of $r = 20\,\mathrm{mm}$, 
width $w = 40\,\mathrm{mm}$, and mass $m = 21.9\,\mathrm{g}$. 
The floor materials differ between the two movies: snow paper 
in Movie S1 and an acrylic plate in Movie S2 (Fig.~\ref{fig:Movie}). 
When the stacked cylinders are compressed from above, the structure 
remains stable on the snow paper (Movie S1). 
In contrast, on the acrylic plate (Movie S2), the lower 
cylinders start to slide, leading to the collapse of the structure.

\end{document}